\documentclass[12pt]{article}

\usepackage{amsmath, amssymb, amsthm}
\usepackage{mathrsfs}
\usepackage{bm}

\usepackage[utf8]{inputenc}
\usepackage[T1]{fontenc}
\usepackage{lmodern}
\usepackage{microtype}
\usepackage{setspace}
\usepackage{hyperref}
\usepackage{epigraph}
\usepackage{algorithm}
\usepackage{algpseudocode}
\usepackage{graphicx}
\usepackage{subcaption}

\usepackage[
  backend=biber,
  style=phys,
  sorting=none,
  biblabel=brackets
]{biblatex}
\title{\vspace{-2cm}Benford's Law from Turing Ensembles \\ and Integer Partitions}

\author{Alexander Kolpakov \\ \href{akolpakov@uaustin.org}{akolpakov@uaustin.org} 
   \and Aidan Rocke \\ \href{rockeaidan@gmail.com}{rockeaidan@gmail.com} }

\date{\today}

\usepackage[a4paper, margin=2.5cm]{geometry}

\usepackage{charter}         

\begin{document}

\maketitle

\begin{abstract}
We develop two complementary generative mechanisms that explain when and why Benford’s first-digit law arises. First, a probabilistic Turing machine (PTM) ensemble induces a geometric law for codelength. Maximizing its entropy under a constraint on halting length yields Benford statistics. This model shows a phase transition with respect to the halt probability. Second, a constrained partition model (Einstein-solid combinatorics) recovers the same logarithmic profile as the maximum-entropy solution under a coarse-grained entropy-rate constraint, clarifying the role of non-ergodicity (ensemble vs.\ trajectory averages). We also perform numerical experiments that corroborate our conclusions. 
\end{abstract}

\section{Introduction}

Benford's law describes the frequency distribution of leading digits in naturally occurring datasets. This logarithmic distribution has been observed in diverse empirical domains, including demographic, economic, and physical measurements \cite{berger2017notices}. Its ubiquity has motivated a wide range of theoretical derivations grounded in scale invariance, base invariance, or entropy maximization \cite{hill1995baseinvariance, Visser2013, Kafri2009EntropyPI}.

While most empirical applications employ base-$10$ (e.g., forensic accounting and scientific measurements), we shall use a more general base-$\Omega$ form. Namely, in base $\Omega$, the probability that a number begins with digit $d$ is given by
\begin{equation}
P(d) = \log_{\Omega}\left(1 + \frac{1}{d} \right), \quad d = 1, 2, \dots, \Omega - 1.
\end{equation}

Our derivations and experiments use base-independent arguments (uniformity of the mantissa on a logarithmic scale), so the conclusions would equally apply for any fixed $\Omega \geq 2$. Classic motivations for base-invariance go back to Newcomb \cite{newcomb1881} and Benford \cite{benford1938} and were formalized by Pinkham \cite{pinkham1961}, Diaconis \cite{diaconis1977}, Berger and Hill \cite{hill1995baseinvariance, bergerhill2011, berger2017notices}. 

Algorithmic perspectives on randomness and encoding have also led to Benford-like statistics. In algorithmic information theory, prefix-free Turing machines are used to define a universal prior over finite binary strings, where the probability of output is proportional to $2^{-\ell(p)}$ for halting programs $p$ \cite{li1997introduction, calude2002random}. Though these models involve deterministic machines, the randomness arises from sampling over programs. The resulting distributions are usually heavy-tailed and emphasize short, low-complexity outputs.

In this work, we adopt a different construction: a probabilistic Turing machine (PTM) with an explicit halting condition, in which binary digits are emitted stochastically until a stop symbol is generated. 

This type of machine, with randomized transitions and halting states, is studied in the theory of randomized computation \cite{arora2009computational}. When constrained by entropy or total codelength, such machines produce ensembles of integer representations whose statistical properties are analyzable using large deviation and maximum entropy principles. We show that under entropy maximization and mild constraints, the digit distribution induced by these PTM ensembles converges to Benford’s law. We validate this interpretation with numerical simulations and compare to empirical datasets with truncated support.

A complementary derivation, based on constrained random partitions in the style of Kafri \cite{Kafri2009EntropyPI}, yields the same logarithmic profile. We interpret this mechanism as a realization of a constrained renormalization flow: while the total digit mass is conserved, and the entropy rate across scales is limited. The logarithmic digit profile then arises as a fixed point of this flow. 

\section{The Information Theory of Benford's Law}

\subsection{A probabilistic model for binary numbers}\label{proba}

We begin by formulating a probabilistic model for generating binary numbers based on a probabilistic Turing machine with a halting condition: this model will serve as the statistical foundation for Benford’s law.

Informally speaking, a (deterministic) Turing machine consists of an infinite tape, a read–write head, and a finite set of states with transition rules; it is strictly more expressive than finite-state machines. A probabilistic Turing machine (PTM) augments transitions with randomness. In our model, each step emits a symbol in $\{0,1,S\}$, and halts when $S$ appears. This yields an analytically tractable stopping-time distribution for output lengths, which is a key ingredient in our maximum-entropy analysis of log-scale statistics. See \cite{li1997introduction, calude2002random, arora2009computational, sipser2012} for detailed background in algorithmic information and randomized computation.

More specifically, let a binary string $\beta$ be produced according to the following rules:
\begin{itemize}
    \item[(1)] For strings of length $\lvert \beta \rvert = 1$, the symbol is `0` with probability $p_0$, or `1` with probability $p_1$;
    \item[(2)] For strings of length $\lvert \beta \rvert > 1$, we set $\beta = ``1"\gamma$, where $\gamma$ is generated by recursively appending `0` with probability $p_0$, `1` with probability $p_1$, or halting with probability $p_S$ by writing the stop symbol $S$.
\end{itemize}
Once $S$ is emitted, generation ceases: the halting symbol $S$ is not included in the output.

We assume the symbols are produced independently and satisfy
\begin{equation}
p_0 + p_1 + p_S = 1.
\end{equation}
We shall consider the symmetric case $p_0 = p_1$ to reflect an unbiased binary generation process.

As we shall see, the halting probability $p_S$ introduces an exponential cutoff in the length distribution of generated strings, giving rise to a characteristic scale. The machine is not prefix-free in the sense of algorithmic coding theory: its output is merely interpreted as the standard binary-encoding. However, this construction serves as a tractable statistical ensemble of integer representations produced by a halting process.

The probability of generating a specific output string depends on both the length and the halting probability, leading to heavy-tailed distributions over binary length and order of magnitude. These features, combined with an entropy-based constraint on the output distribution, naturally lead to the emergence of Benford-like laws, as shown in the next section.

\subsection{Large deviations in Turing ensembles}\label{dev}

We shall study the output of a statistical ensemble of the above PTMs. As we shall see, it exhibits a strong tendency to have large deviations unless the halting probability $p_S$ is high enough. 

Any binary number with $k>1$ bits may be expressed as a code $\beta$ starting with $``1"$ and continued for $k-1$ more random bits until $S$ is generated. Thus, the probability of $\beta$ is given by the geometric distribution 
\begin{equation}
P(k) = (1-p_S)^{k-1}\cdot p_S,
\end{equation}

and the expected length of $\beta$ is given by 
\begin{equation}
\mathbb{E}[|\beta|] = 1 + \mathbb{E}[k-1] = \mathbb{E}[k] = \frac{1}{p_S}.
\end{equation}

The order of magnitude $\Omega(k)$ of a binary number with $k\geq 1$ bits is $2^{k-1}$, if we consider as such the largest value it may take. It follows that 
\begin{equation}
\mathbb{E}[\Omega] = \sum_{k=1}^\infty \Omega(k) \cdot P(k) = p_S \cdot \sum_{k=0}^\infty 2^k \cdot (1-p_S)^k = 
\begin{cases}
\frac{p_S}{2p_S-1}, & \text{if $p_S > \frac{1}{2}$},\\
\infty, & \text{$p_S \leq \frac{1}{2}$}.
\end{cases}
\end{equation}

While $\mathbb{E}[\Omega]$ diverges at the critical point $p_S = \frac{1}{2}$ due to large deviations, we may estimate the typical stopping time and hence the typical order of magnitude using the geometric mean: 
\begin{equation}
\mathbb{E}[\log_2 \Omega] = \mathbb{E}[|\beta|] - 1 \sim \frac{1}{p_S}.
\end{equation}

It follows that the typical number $\beta$ is on the scale of 
\begin{equation}
\chi = 2^{\mathbb{E}[\log_2 \Omega]} \sim 2^{\frac{1}{p_S}}.
\end{equation}
which we call the characteristic scale of the statistical PTM ensemble. In the context of large deviations, $\Omega$ can be thought of as a rate function, while $\chi$ is its entropy \cite{touchette2009}.   

\subsection{From Maximum Entropy to Benford's Law}\label{ben}

We now apply the principle of maximum entropy to constrain the statistical ensemble of PTMs described earlier. Specifically, we seek the parameter values that maximize the entropy of the output distribution, subject to a constraint on total binary length. In the absence of other requirements, this results in uniform sampling over bit lengths \( 1 \leq k \leq N \), i.e., a uniform distribution over logarithmic scale.

Given the cumulative distribution function for halting at length \( k = |\beta| - 1 \),
\begin{equation}
F(k) = 1 - (1 - p_S)^k,
\end{equation}
we impose the macroscopic constraint
\begin{equation}
F(k) \approx \frac{k}{N},
\end{equation}
which enforces approximate uniformity in logarithmic length over the finite support \( 0 \leq k \leq N \). That is, we require the ensemble to be as disordered as possible while remaining scale-bounded.

Since exact uniformity is infeasible under the exponential decay of the geometric distribution, we satisfy this constraint on average. For large \( N \), we approximate the discrete average by an integral:
\begin{equation}
(1-p_S)^N = \frac{1}{N} \sum_{k=1}^N (1-p_S)^N \approx \frac{1}{N} \sum_{k=1}^N \big(1-\frac{k}{N}\big)^{N/k} \rightarrow \int_{0}^1 (1-u)^{\frac{1}{u}} du,
\end{equation}
as $N \gg 1$. 

We define the auxiliary constant
\begin{equation}
\lambda = \int_0^1 (1 - u)^{1/u} du \approx 0.23075\ldots,
\end{equation}
yielding the scaling relation
\begin{equation}
p_S = 1 - \lambda^{1/N},
\end{equation}
which implies \( p_S \to 0 \) as \( N \to \infty \), i.e., vanishing halting probability in the thermodynamic limit.

The constant \( \lambda \) plays a key role in mediating the trade-off between entropy maximization and halting feasibility. It defines the optimal decay rate for the stopping probability \( p_S \) such that the ensemble approximates a uniform distribution in logarithmic scale, while still ensuring termination. Thermodynamically, it acts as an inverse temperature that sets the characteristic length scale of outputs; information-theoretically, it serves as a normalization constant linking entropy constraints with feasible code generation. 

The bit generation probabilities become
\begin{equation}
p_0 = p_1 = \frac{\lambda^{1/N}}{2} = p_{\lambda,N},
\end{equation}
which approach uniformity \( \frac{1}{2} \) as \( N \to \infty \), as expected under entropy maximization.

With these parameters, the cumulative distribution simplifies to
\begin{equation}
F(k) = 1 - \lambda^{k/N} \approx \frac{k}{N} \cdot \ln\left(\frac{1}{\lambda}\right),
\end{equation}
via the Taylor expansion
\begin{equation}
x^{k/N} = \sum_{n=0}^\infty \frac{k^n (\ln x)^n}{n! N^n}.
\end{equation}

Within each fixed length class \( k \), the generation probability is constant:
\begin{equation}\label{eq:inside_order_of_mag}
p(k, \lambda, N) = p_{\lambda,N}^{k-1},
\end{equation}
so each number of given bit length is equally likely, i.e., local uniformity holds inside each order of magnitude.

Since
\begin{equation}\label{eq:orders_of_mag}
P(k) = F(k) - F(k - 1) = \frac{1}{N} \cdot \ln\left(\frac{1}{\lambda}\right),
\end{equation}
the ensemble is approximately uniform across all magnitudes \( 1 \leq k \leq N \), establishing a flat distribution in \( \log_2 X \), for an output \( X \) from the PTM ensemble.

Hence, \( \log_2 X \) is distributed approximately uniformly. In fact, the order of magnitude $k = \lfloor \log_2(X) \rfloor$ is distributed uniformly because of \eqref{eq:orders_of_mag}, and within each order of magnitude $\{ \log_2(X) \}$ tends to the uniform distribution because we have a larger and larger amount of numbers generated each with the same probability \eqref{eq:inside_order_of_mag}. 

This yields that $d \cdot 10^m \leq X < (d + 1) \cdot 10^m$ implies 
\begin{equation}
\log_2 d + m \log_2 10 \leq \log_2 X < \log_2(d + 1) + m \log_2 10,
\end{equation}
so that the probability of first digit \( d \) is proportional to the length of the corresponding logarithmic interval:
\begin{equation}
P(d) \sim \log_2\left(1 + \frac{1}{d}\right).
\end{equation}

After normalization, this reproduces Benford's law. Thus, under a maximum entropy principle constrained by halting length, the PTM ensemble gives rise to a uniform distribution in logarithmic space, from which Benford statistics emerge naturally. Though the argument is heuristic and not fully rigorous, it is supported by both asymptotics and empirical simulations.

\subsection{Numerical experiments}\label{exp}

We perform some experiments to support our conclusions: the corresponding Python code is available on Github \cite{github-benford}. First, we generate binary strings as prescribed by our model, convert them to decimals, and take the leading digit. In each case we generate $10000$ binary strings with $N = 64$ and $N = 128$ (which $N$ is the largest binary length for Benford's law to hold \textit{a priori}, see Section~\ref{ben}) and $\lambda \in \{10^{-12}, 10^{-6}, 0.1, 0.230759776818,$ $0.3, 0.4, 0.5, 0.6, 0.7, 0.8, 0.9\}$. 

\begin{figure}
    \centering
    \begin{subfigure}[t]{0.48\linewidth}
        \centering
        \includegraphics[width=\linewidth]{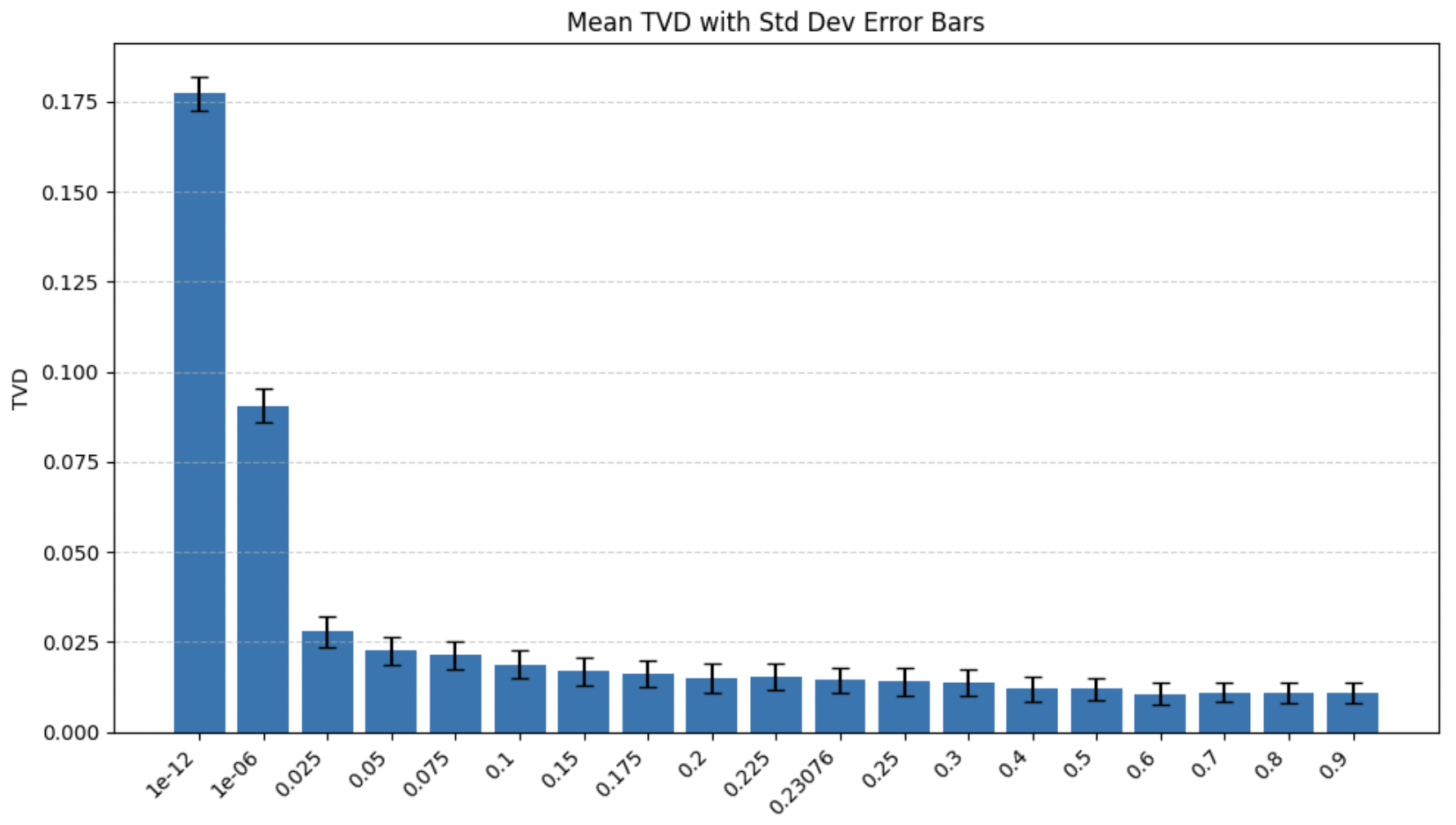}
        \caption{$N=64$}
        \label{fig:lambda-64}
    \end{subfigure}
    \hfill
    \begin{subfigure}[t]{0.48\linewidth}
        \centering
        \includegraphics[width=\linewidth]{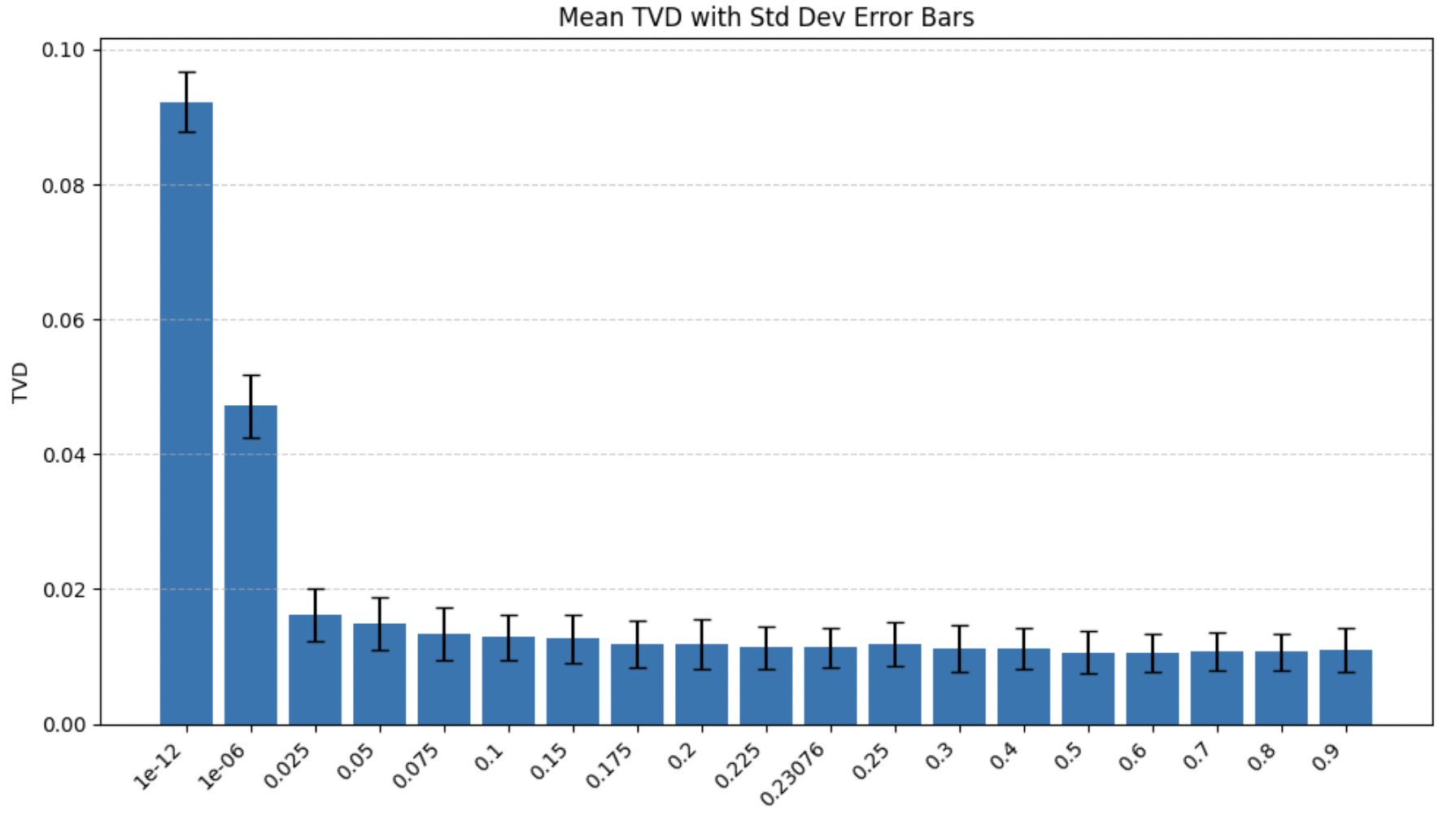}
        \caption{$N=128$}
        \label{fig:lambda-128}
    \end{subfigure}
    \caption{TVD from the model distribution of first significant digits to Benford's law. The model parameter $\lambda$ responsible for the halting probability is on the horizontal axis: $\lambda = 10^{-12}, 10^{-6}, 0.025, 0.05, 0.075, 0.1, 0.15, 0.175, 0.2, 0.23076 (\approx \lambda_*), 0.225, 0.25, 0.3, 0.4, 0.5, 0.6, 0.7, 0.8, 0.9$. Another parameter $N$ is fixed throughout. We use $100$ trials for each value of $\lambda$ to generate the empirical distribution. Error bars are set to $1$ standard deviation.}
    \label{fig:lambda}
\end{figure}

\begin{figure}
    \centering
    \begin{subfigure}[t]{0.48\linewidth}
        \centering
        \includegraphics[width=\linewidth]{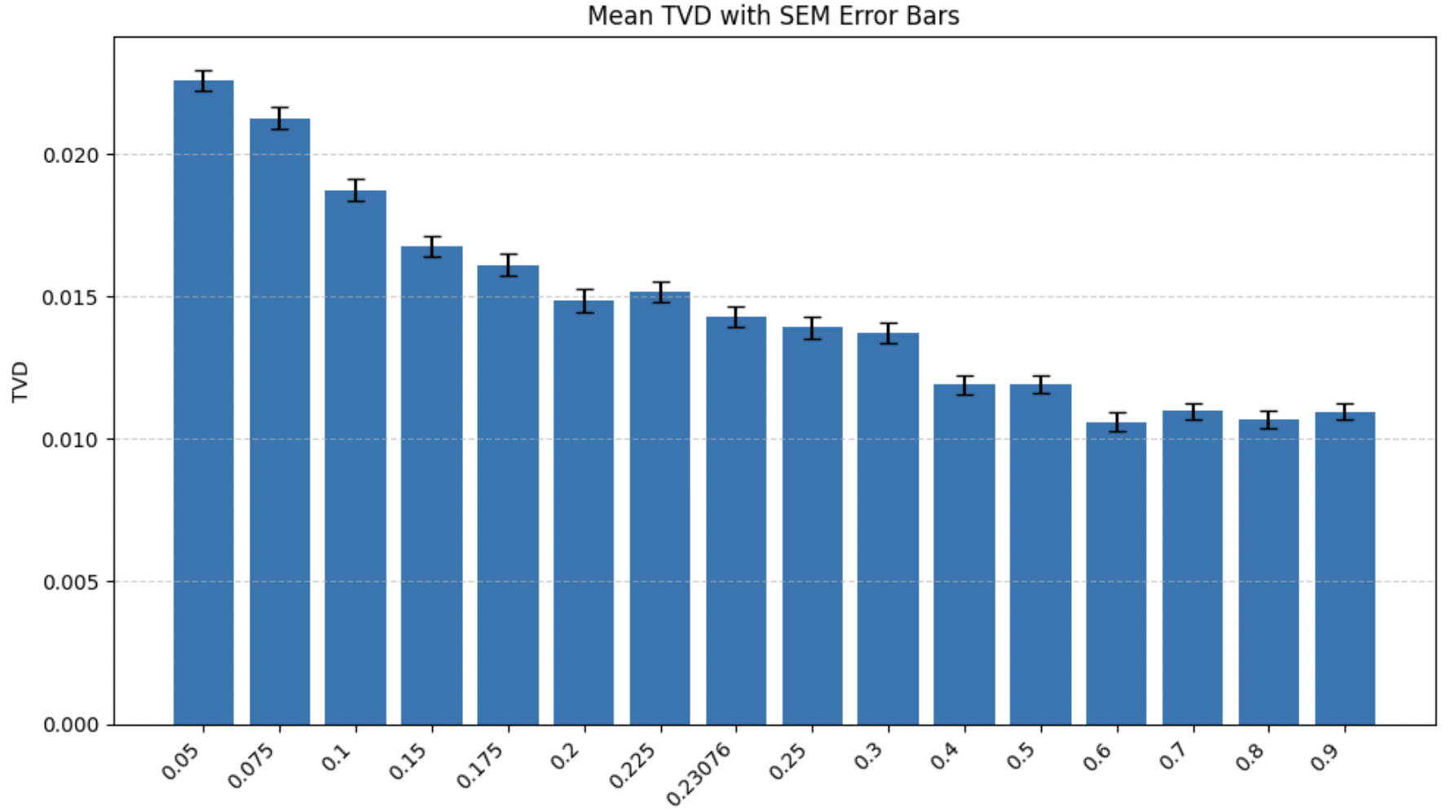}
        \caption{$N=64$}
        \label{fig:lambda-zoom-64}
    \end{subfigure}
    \hfill
    \begin{subfigure}[t]{0.48\linewidth}
        \centering
        \includegraphics[width=\linewidth]{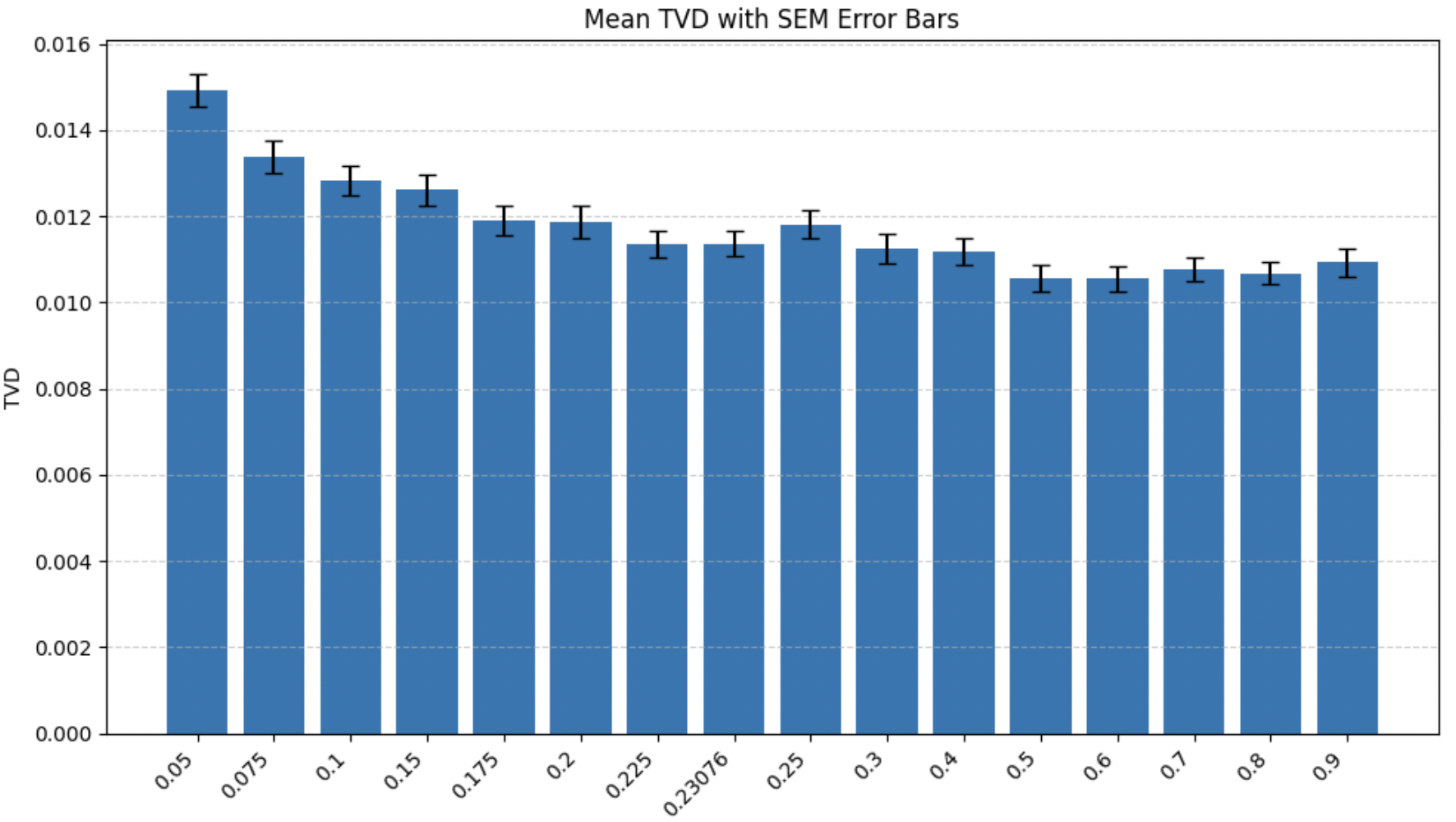}
        \caption{$N=128$}
        \label{fig:lambda-zoom-128}
    \end{subfigure}
    \caption{The same empirically measured TVD zoom around $\lambda_* \approx 0.230759776818$. A visible plateau starts at around $\lambda_*$ when the distribution of the order of magnitude $k = \lfloor \log_2(X) \rfloor$ becomes close to uniform. However, there is a slight downward slope for $N=64$ due to the higher-order effects within each order of magnitude $\{\log_2(X)\}$. No such slope is visible for $N=128$. We use $100$ trials for each value of $\lambda$ to generate the empirical distribution. Error bars are set to $1$ standard error of the mean (SEM).}
    \label{fig:lambda-zoom}
\end{figure}

\newpage

Empirically, we observe that
\[
\lambda_* = \int^1_0 (1-u)^{1/u}\,du \approx 0.230759776818\ldots 
\]
indeed defines a transition threshold, see Figures~\ref{fig:lambda}-\ref{fig:lambda-zoom}. 

The plateau starts at $\lambda = \lambda_*$. For \( \lambda < \lambda_* \), the ensemble fails to match Benford’s law, while for \( \lambda \geq \lambda_* \), the digit distribution converges rapidly to the Benford profile. Indeed, for $\lambda \geq \lambda_*$ we see that the total variation distance (TVD) drops down significantly, after which it has a slight downward slope. More computational experiments are available in the GitHub repository \cite{github-benford}.

Second, we choose the dataset of city populations available from \cite{city-data}. In contrast, this dataset turns out not to follow Benford's law very closely:  the TVD between the empirical distribution and Benford's law in this case is about $0.279$. 

The interesting observation is that were the dataset modeled by our approach, it would have halting probability $p_S \approx 0.054682$ and the average binary length $\approx 18.28$. The maximum binary length of the dataset is $N=25$.

If we use the PTM ensemble with $p_S = 0.054682$, we get a close average binary length, as expected. However, the maximum binary length is much larger: the exact values depend on the trial run, but all are on the order of \emph{hundreds}. In this case, the TVD between the model distribution and Benford's law is about $\approx 0.035$, which is a much closer match (see \cite{github-benford}).

This comparison highlights a key structural insight: empirical datasets may deviate from Benford’s law not due to intrinsic non-conformity, but due to truncation effects in their support. 

The observed discrepancy between the city population data and the Benford distribution likely stems from the finite-size cutoff at $N = 25$, which suppresses high-order magnitudes and introduces statistical bias in the leading digit distribution. When we extend the support of the PTM ensemble to $N > 100$ while preserving the empirical average bit-length via $p_S = 0.054682$, the digit distribution converges sharply to Benford’s law. 

Our observation suggests that matching only first-order moments (e.g., mean codelength) is insufficient: the tail behavior of the ensemble (determined by the halting probability) plays an important role in recovering scale-invariant statistics.

\section{Non--ergodicity of large integer partitions}\label{kafri}

The following analysis is based on Kafri's statistical mechanical model for Benford’s law~\cite{Kafri2009EntropyPI}, which approaches the digit distribution problem through a statistical mechanics framework. Unlike our earlier model based on a PTM ensemble, Kafri models digit sequences as components of a constrained random vector. This links Kafri's model to the classical theory of integer partitions \cite{andrews1998, hardy-ramanujan1918, erdos-lehner1941}. 

Here, we reinterpret his framework in light of non-ergodic dynamics and constrained entropy maximization. We show that the non-coincidence of spatial and temporal digit averages directly yields a logarithmic bias, offering an alternative yet consistent explanation of Benford's law.

An $N$-digit number in base $\Omega \in \mathbb{N}$ may be modeled as a random vector with $N$ components where each component has value between $0$ and $\Omega - 1$. Let us consider the function $\phi(\sigma)$ that counts the number of components of $\vec{X}_N$ that equal $\sigma$. Also, we have
\begin{equation}
    \frac{1}{N} \sum^{\Omega-1}_{\sigma=0} \phi(\sigma) = 1, 
\end{equation} 
which means that $\phi$ plays the role of a counting measure. 

Let the ``average digit'' of  $\vec{X}_N$ be
\begin{equation}
    \mu = \frac{1}{N} \sum^{\Omega-1}_{i=0} \sigma \cdot \phi(\sigma). 
\end{equation} 

At time step \( 0 \leq i \leq N \), we consider the coarse-scale behavior of the system under a constraint on the total digit sum \( \mathcal{E} = N \mu \). To do so, we introduce a rescaled variable \( t = \frac{i \mu}{N} \), which traverses the digit mass from $0$ to \( \mu \). 

We assume that, under constrained entropy maximization, the local digit content at scale \( t \) is described by a continuous density \( \phi(t) \), representing the digit weight per unit mass at effective resolution \( t \). The function \( \phi \) may be viewed as a generalized or interpolated histogram over digit values.

We then evaluate the total contribution of digit content across scales by computing
\[
\frac{1}{N} \sum_{i=0}^N \mu \cdot \phi\left( \frac{i \mu}{N} \right) \longrightarrow \int_0^\mu \phi(t)\,dt.
\]

This integral may be interpreted as a renormalized observable: it aggregates digit contributions across coarse-grained scales, under the constraint that the final state has mean digit \( \mu \). Although no degrees of freedom are eliminated, the use of progressively scaled arguments \( t = \frac{i \mu}{N} \) defines a resolution flow over the digit space.

If we assume that $\vec{X}_N$ is generated in ergodic manner, then the spatial and temporal averages of our dynamical system would coincide, as ergodicity would imply via Birkhoff's theorem. This assumption results in the trivial solution $\phi(\mu) = 1$. Thus, we must essentially employ the \textit{non-ergodicity} of the process. 

Let us constrain first the spatial average of our sequence:
\begin{equation}
    \sum^{\Omega-1}_{i=0} \sigma \cdot \phi(\sigma) = \mu N = \mathcal{E} = const.
\end{equation}

To estimate the number of vectors $\vec{X}_N$ with magnitude constraint $\mathcal{E}$, it suffices to consider the combinatorial expression: 
\begin{equation}
\Lambda(N, \mathcal{E}) = \frac{(N+\mathcal{E}-1)!}{(N-1)!\; \mathcal{E}!}.
\end{equation}

By using Stirling's approximation, we find the entropy of the partition ensemble: 
\begin{equation}\label{entropy:simplified}
\ln \Lambda \approx N \big((1+\mu)\cdot \ln (1+\mu) - \mu \cdot \ln \mu \big).
\end{equation}

However, the temporal entropy should also be constrained in some way. Here we posit that we can only constrain the time average of digits. This means that the entropy rate is limited: the generation of each next digit has bounded entropy, while the whole process is open-ended and unbounded in time. 

Thus, we set
\begin{equation}
    \int^\mu_0 \phi(t) dt = \varepsilon = const.
\end{equation}

In this sense, \( \phi(t) \) may be understood as the (scale-dependent, in general) digit density, and the integral constraint represents a renormalization condition on the entropy rate. 

Now, in order to find $\phi(\mu)$ that maximizes entropy we specify a Lagrangian in terms of the entropy $\ln \Lambda (\mu)$ and the entropy constraint:

\begin{equation}
\mathcal{L}(\mu) = \ln \Lambda - \beta \cdot \left(\int^\mu_0 \phi(t) dt - \varepsilon\right).
\end{equation}

By using \eqref{entropy:simplified} we find
\begin{equation}
\frac{\partial \mathcal{L}}{\partial \mu} = 0 \implies \phi(\mu) = \frac{N}{\beta}\ln \big(1+ \frac{1}{\mu}\big),
\end{equation}

as well as
\begin{equation}
\langle \phi \rangle = \frac{1}{N} \sum_{\mu=1}^{\Omega-1} \phi(\mu) = \frac{1}{\beta} \ln \Omega.
\end{equation} 

As the probability of the first digit ought to be normalized in order to be dimensionless, we find 
\begin{equation}
P(n) = \frac{\phi(n)}{\sum_{n=1}^{\Omega-1} \phi(n)} = \frac{\ln \left(1+\frac{1}{n}\right)}{\ln \Omega} = \log_{\Omega} \left(1+\frac{1}{n}\right),
\end{equation}
which is equivalent to Benford's law. 

Going back to our PTM model, let us note that it has an absorbing halt state, so a single run explores only a fraction of the tree-like configuration space; thus time-averages along trajectories need not match ensemble-averages over completed outputs. This aligns with Kafri’s insight that Benford-like behavior reflects an imbalance between the volume of admissible configurations (state-space) and the accessible temporal trajectories. 

Therefore, we carry out Jaynes' Maximum Entropy analysis \cite{jaynes1957a} as an inference principle over the ensemble of \emph{halting} PTMs with completed outputs by maximizing Shannon's entropy under macroscopic constraints.

\section*{Acknowledgments} 

This material is based upon work supported by the Google~Cloud Research Award number GCP19980904. 

A.K. was also supported by the Wolfram Institute for Computational Foundations of Science, and by the John Templeton Foundation. 

\printbibliography

\end{document}